\documentclass{moriond}

\usepackage{amsmath}

\def\be{\begin{equation}}
\def\ee{\end{equation}}
\def\bea{\begin{eqnarray}}
\def\eea{\end{eqnarray}}

\newcommand{\Photo}{}

\begin{document}
\vspace*{4cm}
\title{About non-classical gravitational states}

\author{ Federico Piazza }

\address{Aix Marseille Univ, Universit\'{e} de Toulon, CNRS, CPT, Marseille, France}

\maketitle
\abstracts{Take the gravitational field, a number of macroscopic observers and put them in a ``highly-quantum" state, made by a coherent superposition of different classical configurations. How would that look like? I try to study this system by introducing a spatial distance operator among the observers and looking at the properties of its expectation values.}

\section{Introduction}

By a \emph{classical state} here I mean a state whose wave-function is peaked around  
one solution of the classical equations of motion. The situation is famously illustrated by coherent states, centered around a classical trajectory and just ``furbished" with  the minimal uncertainty allowed by the Heisenberg principle. 
The double-well potential offers an equally known example of \emph{deviations} from classicality. Its ground state does not correspond to a classical configuration sitting on either of the minima, but is a coherent superposition of the two. As nicely explained by Coleman~\cite{Coleman:1985rnk}, this  generalizes  to periodic potentials and, most notably, to gauge theories. The ground state of a non-Abelian Yang-Mills theory \emph{is not} just $A^\mu = 0$ ``furbished" with the minimal amount of quantum indeterminacy. The \emph{$\theta$-vacua} of such theories are coherent superpositions of topologically inequivalent configurations. When time-dependent or transient, deviations from classicality are normally associated with tunneling.

Hovering over the latter examples are \emph{Euclidean} path integrals. 
Broadly speaking, when the relevant classical saddles live in Euclidean spacetime and are non-trivial, in the sense that they interpolate different classical configurations, this often signals a departure from classicality.

Recently, gravitational Euclidean methods have played an important role in addressing aspects of the black hole information puzzle~\cite{Almheiri:2020cfm}. Non-trivial Euclidean saddles connecting different boundaries account for the decrease of entropy in the late-Hawking radiation, as expected if the whole system evolves unitarily. 
It is tempting to wonder if  \emph{gravitational} deviations from classicality at low energy are at play in this context. Perhaps related to all this, gravitational instantons/wormholes have been advocated as a non perturbative mechanism to solve the cosmological constant problem~\cite{Hebecker:2018ofv}. 
But what would it mean for the gravitational field to be in a coherent superposition of different classical metrics? What would an observer living in such a spacetime experience? 

In this talk I report on recent little advances in understanding these issues~\cite{Piazza:2021ojr}.  I will show that, perhaps counterintuitively, a highly quantum gravitational state does not seem to imply anything exotic, \emph{locally}. However, anomalous geometrical effects can build up at large separations.

\section{A fluid (or a solid) of observers}

I am interested in low energy phenomena, governed by Einstein gravity. 
In the standard canonical theory~\cite{DeWitt:1967yk} the spatial metric $h_{ij}$ is the relevant dynamical variable. Physical states satisfy Hamiltonian and momentum constraints. 
The idea is then to consider a highly quantum such state $\Psi[h_{ij}(x^k)]$ and try to understand it ``operationally", i.e. in terms of the mutual observations of a bunch of observers living therein. 
One should thus introduce additional fields in the low energy theory to model the observers and their mutual interactions. 

My choice of observers is that of a continuum ``fluid" of them. The effective field theory description of a fluid or a solid~\cite{Dubovsky:2005xd} involves three scalar fields $X^I$, $I= 1,2,3$. Each triplet of values  labels and ``follows in time" an infinitesimal volume of the medium. Equivalently, each triplet is the \emph{name} of an observer.  One can arrange the Lagrangian in such a way that  $X^I = const.$ is a geodesic on any classical solution. In this case we are dealing with a non-relativistic fluid. However, for the moment, I will maintain agnostic about the dynamics of the medium.

The three fields $X^I $ effectively serve as a reference system, to which I will add a fourth ``time" field $T$. Dynamical reference systems have a long story as a mean to define gauge invariant observables~\cite{Rovelli:1990pi} and dress local operators~\cite{Marolf:2015jha} in quantum gravity. Ideally one would like to add also a light field  to model signal exchanges among the observers. 
A more economic strategy is to focus  directly on spacetime \emph{distances} as a proxy for all such mutual relations.\footnote{E.g. if a light ray connects events $A$ and $B$ then $A$ and $B$ are at null Lorentzian distance. See also~\cite{Witten:2019qhl} on this.} 

In the canonical theory the fields introduced above are commuting, and thus can be used as independent variables in the wave-function. It is known that $ \Psi[h_{ij} (x^k), \, X^I(x^k), \, T(x^k)] $, despite depending of spatial quantities, portraits a portion of space\emph{time}. The point is that the Hamiltonian constraint makes $\Psi$ invariant under time-diffeomorphisms.  
However, my very limited task here is to define \emph{spatial distances between the observers along a given time slice}, which can be made gauge-invariant with the aid of the reference fields $X^I$ and $T$.

\section{Spatial-distance operator}

I will use the field $T$ to define the \emph{conditional probability amplitude} that all remaining fields be in some configuration \emph{given that} the value of $T(x^k)$ is a constant, $T_0$,
\begin{equation}
\tilde \Psi[h_{ij} (x^k),\, X^I(x^k);T_0] \ \equiv\ {\cal N}(T_0)  \Psi[h_{ij} (x^k), \, X^I(x^k),  \, T(x^k) =T_0] . \
\end{equation}
 This is a portrait of the system ``at a given time". The  $\tilde \Psi$-functionals only depend on $h_{ij}$ and $X^I$ and must satisfy the momentum constraint. A convenient (overcomplete) basis for the corresponding Hilbert space are the common eigenvectors of such commuting variables, $|h_{ij} (x^k), \, X^I(x^k)\rangle$. Each element of this basis represents a very classical three-dimensional geometry with the positions of the observers perfectly specified on it. On $|h_{ij} (x^k), \, X^I(x^k)\rangle$ it is thus possible to calculate the distance between any chosen pair of observers.

So I define a spatial distance operator between $\vec X$ and $\vec Y$, $\hat d(\vec X, \vec Y)$, by declaring that it is diagonal on $|h_{ij} (x^k), \, X^I(x^k)\rangle$, 
\begin{equation}
\hat d(\vec X, \vec Y)\  |h_{ij} (x^k), \, X^I(x^k)\rangle \ = \  d(\vec X, \vec Y)_{\{h,  X^I\}} \ |h_{ij} (x^k), \, X^I(x^k) \rangle\, .
\end{equation}
In the above equation the two triplets $\vec X$ and $\vec Y$ are the  names of the two observers and the eigenvalue $d(\vec X, \vec Y)_{\{h,  X^I\}}$ is their distance on the corresponding eigenstate.  In practice, as usual, one needs to find the geodesic between $\vec X$ and $\vec Y$ and integrate the spatial line element along it.  The subscript ${\{h,  X^I\}}$ is there to remind us that this procedure depends on the geometry and on the fields' configuration on the eigenstate $|h_{ij} (x^k), \, X^I(x^k)\rangle$.

\section{Expectation values}

On a generic state $\tilde \Psi$, expectation values of $\hat d(\vec X, \vec Y)$ are calculated by expanding $\tilde \Psi$ on the classical eigenbasis, 
\begin{equation} \label{average}
\left\langle  \hat d(\vec X, \vec Y)^2  \right  \rangle =  \sum_I \int \widetilde{{\cal D}  h}\,  {\cal D} X^I 
\left|\langle \tilde  \Psi | h, X^I\rangle\right|^2 \ \left(d(\vec X, \vec Y)_{\{ h,  X^I\}}\right)^2 \, .
 \end{equation}
 \begin{figure}[h]
\vspace{-.9cm}
\begin{center}
\includegraphics[width=6 cm]{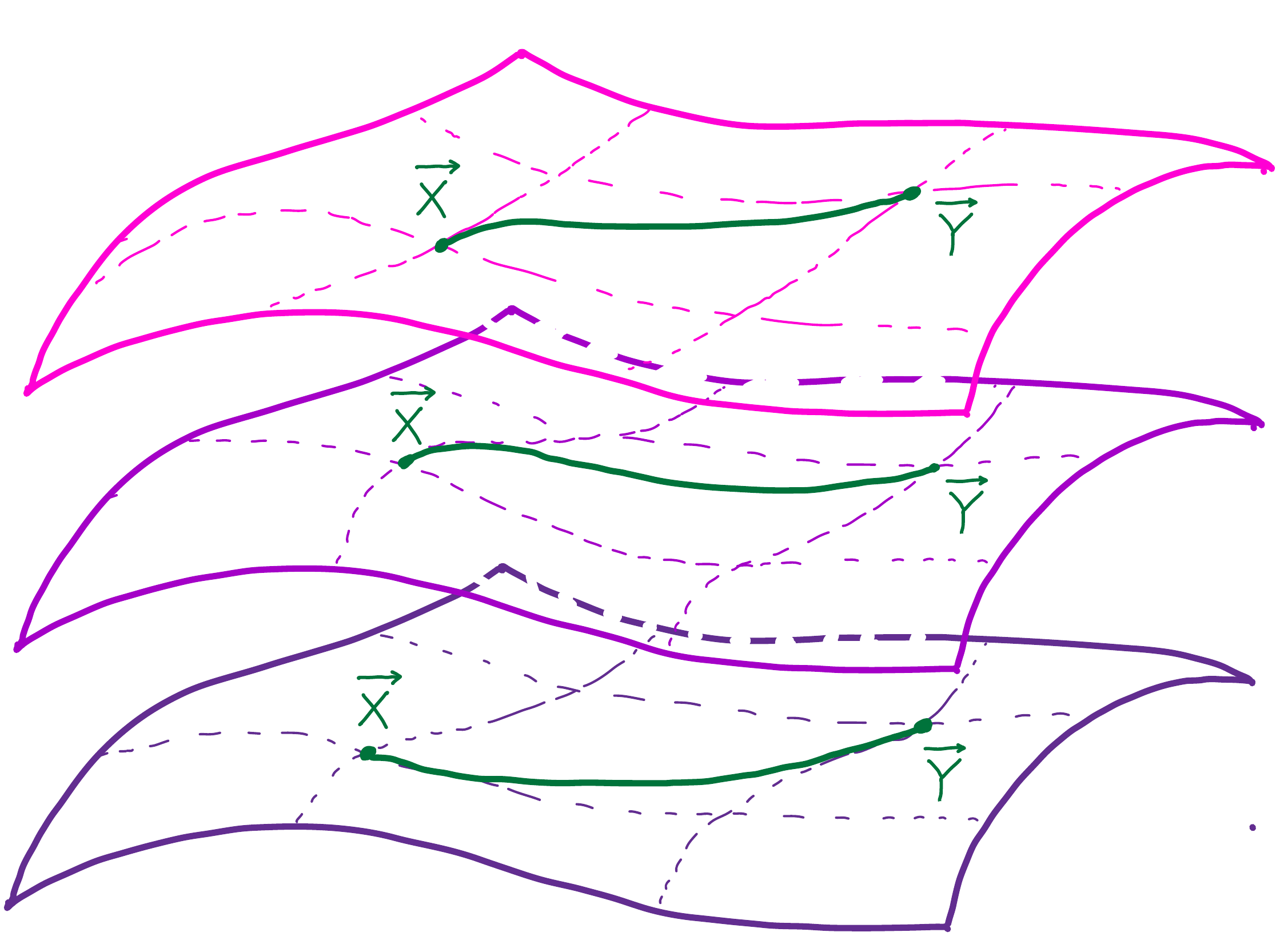}
\end{center}
\caption{Every classical configuration contributes to the distance expectation value, weighted by the amplitude probability $\tilde \Psi$. On each configuration the distance between $\vec X$ and $\vec Y$ is a well-defined calculable quantity.} \vspace{-.3cm}
\end{figure}

 Because of gauge redundancy, different metric configurations  represent the same geometry.  The tilde on ${\cal D}  h$  means that we should  pick  just one representative for each gauge orbit. 
 
In~(\ref{average}) I have evaluated the average of $\hat d^2$ because it has better analytic properties than that of $\hat d$ and is more naturally generalizable to the Lorentzian case: Lorentzian distances can be either positive or imaginary and averaging over them could be meaningless.   The distances \emph{squared}, instead, are always real and thus average to a positive, negative or null value.

The observers' labels $\vec X$ define a coordinate system on a three-dimensional manifold which is very naturally equipped with the distance
\begin{equation}\label{distance}
\overline{d(\vec X, \vec Y)} \equiv \sqrt{\left\langle \hat d(\vec X,\vec Y)^2\right\rangle}\, .
\end{equation}
Is~(\ref{distance}) itself a \emph{Riemannian} distance?  In other words, is there an ``average line element" on the manifold that reproduces (\ref{distance}) for each pair of points? It will be sufficient to show a counterexample in order to give a negative answer to this question. 
\section{Average distances are non-additive}
Consider a superposition of two classical eigenvectors, $
|\tilde \Psi \rangle = (|\tilde \Psi_1 \rangle + |\tilde \Psi_2 \rangle)/\sqrt{2}\, .$
The system is one-dimensional and the classical states $|\tilde \Psi_1 \rangle$ and  $|\tilde \Psi_2 \rangle$ are depicted in the figure below. Only four observers are picked out in the figure: $A$, $B$, $C$, and $D$. 
\begin{figure}[h]
  \centering \vspace{-.3cm}
  \includegraphics[width = .6\linewidth]{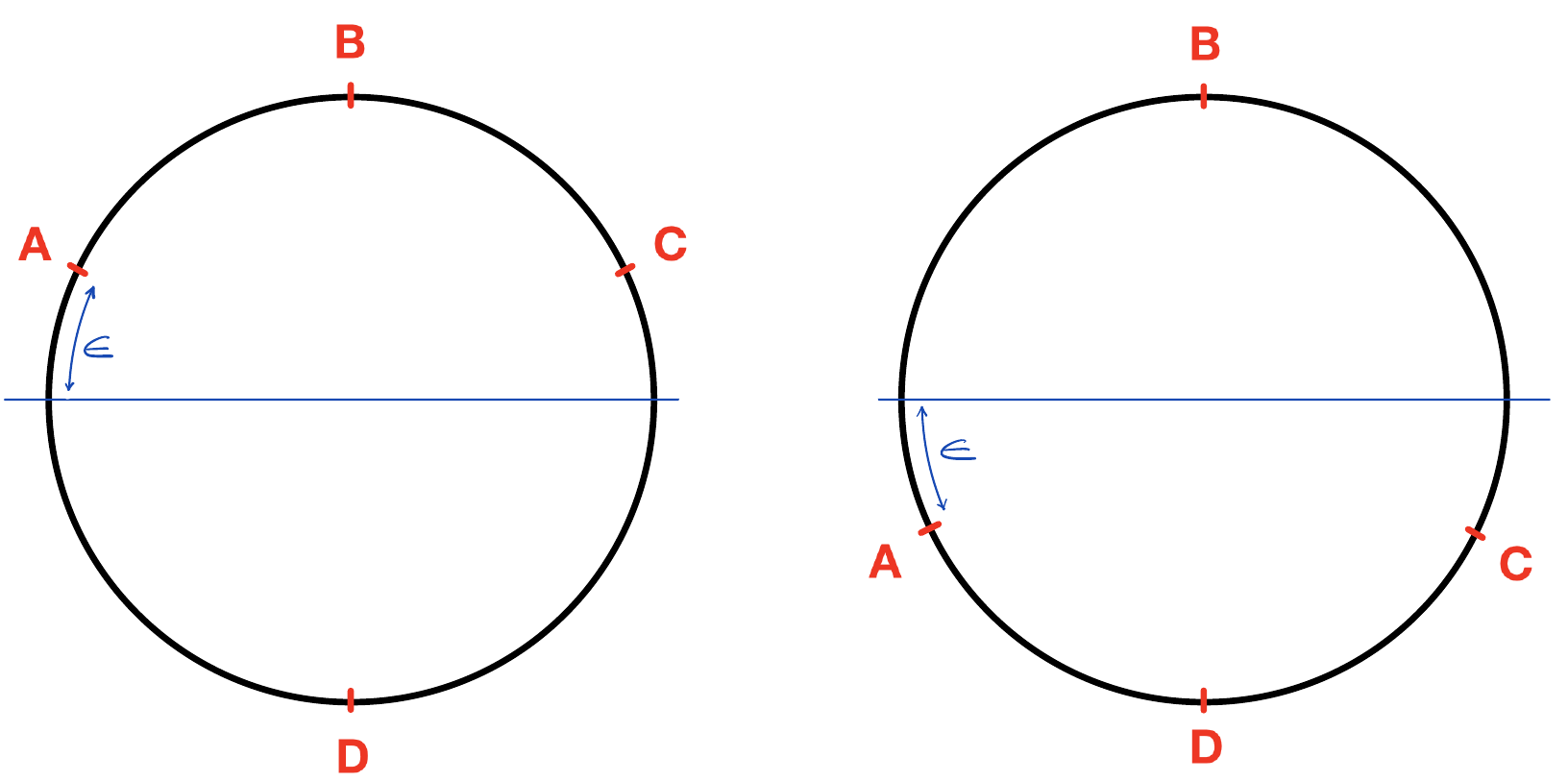} \\[.1cm]
   \large $\displaystyle \ |\tilde \Psi_{1} \rangle  \ \ \qquad \quad \qquad \qquad \qquad  \qquad \qquad |\tilde \Psi_{2} \rangle   $
  \end{figure}
  
To first order in $\epsilon$, 
\begin{align}
\overline{d(A,B) }  = \overline{d(B,C) } = \overline{ d(A,D) } = \overline{ d(C,D) } = \pi/2 \, , \quad \ 
\overline{ d(B,D) }  =  \pi \, ,  \quad\  \ 
\overline{ d(A,C) }  = \pi - 2 \epsilon \, .
\end{align}
It is clear from this example that average distances \emph{are not} Riemannian distances. In particular, even in one dimension they fail to add up: $\overline{d(A,B) }+ \overline{d(B,C) }\neq \overline{d(A,C) }$.

This anomalous behavior, in general, builds up at large separation and is reminiscent of some IR non-local effect found in double holography~\cite{Omiya:2021olc}. On the opposite, in the neighbors of each observer nothing particularly exotic happens. There is~\cite{Piazza:2021ojr} an average metric ${\cal G}_{IJ}$ such that 
\begin{equation} \label{EP}
\overline{d(0,\vec X)} \ =  \ \sqrt{{\cal G}_{IJ}  \ X^I X^J }\ + \ {\cal O}(X)^3\, ,
\end{equation}
where the labels $\vec X$ have been used directly as coordinates and one of the two observers has been located at the origin. Overall, average distances behave like \emph{chord distances} of embedded manifolds: they have a standard locally flat limit but fail to be additive at large separation~\cite{Piazza:2021ojr}.

\section{Conclusions}
I have tried to study departures from classicality in gravity with the aid of certain auxiliary scalar fields that serve as a reference frame of ``observers". 
The expectation values of their mutual distances are a good diagnostic of the quantum-ness of the state. Most strikingly, average distances can grow in some non-additive way while maintaining a regular behavior at small separations. The physical implications of this effect are far from obvious---space-like distances do not have direct operational meaning anyway. However, if some of this non-additivity leaks along the time direction, this could have interesting implications e.g. on causality~\cite{Piazza:2021ojr}~\cite{Omiya:2021olc}.

My considerations are purely kinematical in the sense that, from the onset, I consider states that already deviate from classicality. 
There is no firm indication, at present, about when and how a low-energy gravitational system could find itself in such a highly quantum state. However, it is not difficult to see that the  non-additive effect just described is always going to be there, to some extent. Perhaps it is just  negligible in most realistic situations.  
It would be interesting to quantify this effect, say, in a Hartle-Hawking state. It would be nice also to extend this study to four-dimensional spacetime distances, which have more direct operational meaning.

\end{document}